\documentclass[9pt, aps, nofootinbib, prd, onecolumn, notitlepage]{revtex4-1}

\usepackage{amsmath}
\usepackage{amssymb}
\usepackage{bm}% bold math
\usepackage[usenames,dvipsnames]{color}
\usepackage{dcolumn}% Align table columns on decimal point
\usepackage{epsfig}
\usepackage{graphicx}% Include figure files
\usepackage{newtxmath, newtxtext}
\usepackage[pagebackref=false, colorlinks=true]{hyperref}
\usepackage{subfigure}
\usepackage{mathrsfs}

\definecolor{reddish}{rgb}{0.7,0.2,0.0}  
\definecolor{blueish}{rgb}{0.1,0.1,1}
\hypersetup{
linkcolor=reddish,         % color of internal links
citecolor=blueish,           % color of links to bibliography
filecolor=blue,              % color of file links
urlcolor=magenta}       % color of the url

\begin{document}

\title{Comment on the Analytical Bounds in the Rezzolla-Zhidenko
  Parametrization}

\author{Prashant Kocherlakota$^{\text{1}}$ and Luciano Rezzolla$^{\text{1,2,3}}$}

\affiliation{$^{\text{1}}$Institut f{\"u}r Theoretische Physik,
  Goethe-Universit{\"a}t, Max-von-Laue-Str. 1, 60438 Frankfurt, Germany}
\affiliation{$^{\text{2}}$Frankfurt Institute for Advanced Studies,
  Ruth-Moufang-Str. 1, 60438 Frankfurt, Germany}
\affiliation{$^{\text{3}}$School of Mathematics, Trinity College, Dublin
  2, Ireland}

\begin{abstract}
In this short note, we briefly comment on the analytical bounds that must
be imposed on the parameter space of the Rezzolla-Zhidenko (RZ)
metric-parametrization approach introduced in
Ref.~\cite{Rezzolla+2014}. We hope this will clarify some of the
confusion recently emerged on this issue \cite{Heumann+2022}.
\end{abstract}

\maketitle

\textit{Introduction.} The Rezzolla-Zhidenko (RZ) metric parametrization
approach was introduced in Ref.~\cite{Rezzolla+2014} to characterize
efficiently and accurately the exterior horizon geometry of
asymptotically-flat, spherically-symmetric and static black-hole (BH)
spacetimes. This approach has been employed successfully to model a large
number of generic BH spacetimes\footnote{As noted in
Ref. \cite{Kocherlakota+2020}, the RZ parametrization can trivially be
employed also for spherically symmetric static spacetimes that do not
contain BHs but other compact objects, including nonsingular and singular
black holes, boson stars, and naked singularities.} in terms of a series
of coefficients $\{\epsilon, a_i, b_i\}$ and by Pad\'e approximants in
the form of continued fractions (see Ref. \cite{Kocherlakota+2020} and
references therein); the approach has also been extended to stationary BH
spacetimes in Ref. \cite{Konoplya+2016}.

\medskip
\textit{Comment.} In the RZ-parametrization approach, the radial
coordinate $r=r_0$ must, \textit{by construction}, correspond to the
location of the outermost Killing horizon. We recall that spherically
symmetric BH spacetimes admit a time-translation Killing vector
field $\boldsymbol{K}$ which, in the static ($\boldsymbol{K} \wedge 
\mathrm{\mathbf{d}}\boldsymbol{K} = \mathbf{0}$) exterior horizon geometry, can be 
identified to be $\boldsymbol{K} =
\boldsymbol{\partial_t}$. $\boldsymbol{K}$ is asymptotically timelike,
i.e., $g_{tt} < 0$, and becomes null on Killing horizons located at
$r=r_{\mathrm{kh};k}$, i.e., $g_{tt}(r_{\mathrm{kh};k}) = 0$ (with $k$
indexing the roots of $g_{tt}$). As per the definition above, if $r_0$ is
the largest, real, positive root, i.e., $r_0 =
\mathrm{max}\{r_{\mathrm{kh};k},~\forall~k\} \in \mathbb{R}^+$, then
clearly for $\boldsymbol{K}$ to remain timelike on the exterior horizon
geometry, $N^2(r):= - g_{tt}$ must be positive-definite on $r >
r_0$. Therefore, and as already noted in Ref.~\cite{Rezzolla+2014,
  Kocherlakota+2020, Kocherlakota+2022, EHTC+2022f}, this requires that
the following condition must be met
\begin{align}
  \label{eq:A_Condition}
  N^2 =: x A(x, \epsilon, a_i) > 0\,,
  \qquad \mathrm{for}~0\leq x \leq 1\,,
\end{align}
where $x: = 1-r_0/r$ is a compactified coordinate such that $x=0$
corresponds to the location of the event horizon ($N^2(x)=0$) and $x=1$
to spatial infinity (we recall that $r_0 := 2M/(1+\epsilon)$, where $M$
is the BH mass). As pointed out already in 2014 (see Eq. (5) of
Ref.~\cite{Rezzolla+2014}), Eq.~\eqref{eq:A_Condition} partitions the
relevant parameter space into mathematically ``allowed'' and
``disallowed'' regions, and its impact on the relevant parameter spaces
for various slices of the general RZ metric has \textit{already} been
reported in Table 2 of Ref.~\cite{Kocherlakota+2022}, but also in
Figs.~3-5 of Ref.~\cite{Kocherlakota+2022}, where the regions in which
$r=r_0$ does not locate the outermost Killing horizon are shown as gray
regions.

Recently, Ref.~\cite{Heumann+2022} has considered a two-parameter
realization of the RZ approach, that is, it has employed the space of
paramters $(r_0, a_1)$, or equivalently $(\epsilon, a_1)$, to express the
$g_{tt}$ metric function of a generic BH spacetime and solved for all
possible roots of $N^2(r, r_0, a_1)=0$, which corresponds to a quartic
equation in this specific case. Finding that $r=r_0$ does not locate the
outermost Killing horizon for arbitrary choices of the RZ parameters,
Ref. \cite{Heumann+2022} partitions the relevant $(r_0, a_1)$ RZ
parameter space into regions where $r=r_0$ does in fact locate the
outermost Killing horizon and where it does not, essentially re-iterating
the computation reported in Ref.~\cite{Kocherlakota+2022}. Indeed,
Eqs. (25) and (26) of Ref.~\cite{Heumann+2022} are not new and have
already been reported respectively in the seventh and eight rows of Table
2 of Ref.~\cite{Kocherlakota+2022} (i.e., the case $\mathscr{M}(\epsilon,
a_1)$); the same table also provides the ranges for other choices of the
parametrization.

\medskip
\textit{Conclusion.} Contrary to the recent claim in
Ref. \cite{Heumann+2022}, the horizon radius in the RZ parameterization
does not change discontinuously. This can happen only if the
parameterization is performed incorrectly outside of the region of
validity required by Eq.~\eqref{eq:A_Condition}. Such conditions were
\textit{already} discussed and published in Refs. \cite{Rezzolla+2014,
  Kocherlakota+2020, Kocherlakota+2022, EHTC+2022f}. Finally, we note
that to ensure that the Ricci and Kretschmann scalars corresponding to
the RZ metric do not diverge outside or on the horizon, the metric
function $B^2 (r) := g_{rr}(r) N^2(r)$ must generically be nonvanishing
for $r \geq r_0$. This represents another nontrivial constraint on the
$b_i$ coefficients of the RZ parameter space.

\bigskip
\noindent We thank Alexander Zhidenko for useful comments. Support comes
from the ERC Advanced Grant ``JETSET: Launching, propagation and emission
of relativistic jets from binary mergers and across mass scales'' (Grant
No. 884631).

\bigskip
\bigskip
\bigskip
\bigskip

%%%%%%%%%%%%%%%%%%%%%%%%%
%%%%%%%%%%%%%%%%%%%%%%%%%

%\bibliographystyle{plain}
\bibliography{BIB_Comment_on_Parametrized_Metrics}

%%%%%%%%%%%%%%%%%%%%%%%%%
%%%%%%%%%%%%%%%%%%%%%%%%%

\end{document}